\documentclass{amsart}
\usepackage{amssymb}

\makeatletter

 \let\paragraphname\@empty  

 \newcommand{\nn}{\notag}
\newcommand{\customqed}[1]{{\renewcommand{\qedsymbol}{#1}\qed}}
\newcommand{\varqed}{\customqed{\hbox{$\diamondsuit$}}}

 \theoremstyle{plain}

 \newtheorem{lemma}{Lemma}[section]

 \newtheorem{theorem}[lemma]{Theorem}
 
 \newtheorem{fact}[lemma]{Fact}

 \theoremstyle{definition}

 \theoremstyle{remark}

 \newtheorem{Condition}[lemma]{Condition}

 \newenvironment{condition}{%
   \begin{Condition}}{\varqed\end{Condition}}

 \newtheorem{Example}[lemma]{Example}
 \newtheorem{Examples}[lemma]{Examples}

 \newtheorem{Remark}[lemma]{Remark}
 \newtheorem{Remarks}[lemma]{Remarks}

 \newcommand\ol[1]{\overline{#1}}

 \newcommand \ag{\alpha}
 \newcommand \bg{\beta}

 \newcommand \rg{\varrho}

 \newcommand{\thg}{\vartheta}
 \newcommand \Thg{\Theta}
 \newcommand \tg{\tau}    

 \newcommand{\bH}{\mathbf{H}}

 \newcommand{\bS}{\mathbf{S}}

 \newcommand{\bW}{\mathbf{W}}

 \newcommand{\bZ}{\mathbf{Z}}

 \newcommand{\cI}{\mathcal{I}}

 \newcommand\setof[1]{\mathopen\{\,#1\,\mathclose\}}

 \newcommand\cei[1]{{\lceil #1\rceil}}

 \newcommand \such {:}

 \newcommand\Prob{\mathop{\operator@font Prob}}
 \newcommand\Maj{\mathop{\operator@font Maj}}
 \newcommand\nor{\mathbin{\operator@font NOR}}

  \renewcommand{\le}{\leqslant}
  \renewcommand{\ge}{\geqslant}

 \newcommand {\sbs}{\subset}

 \newcommand {\xcpt}{\mathbin{\raise0.15ex\hbox{$\smallsetminus$}}}

 \newcommand\for{\text{ for }}

 \newcommand{\HyphConv}{\expandafter\HyphConvFirst\HyphConvDelim}
 \newcommand{\HyphConvDelim}[1]{#1\%}
 \newcommand{\HyphConvFirst}{\futurelet\tempa\HyphConvCond}
 \newcommand{\HyphConvCond}{%
    \def\tempb{\HyphConvSubst\tempa}%
    \if\tempa\%\let\tempb\HyphConvStop
    \else\if\tempa-\let\tempa\_\fi
    \fi
    \tempb
  }
 \newcommand{\HyphConvStop}[1]{}
 \newcommand{\HyphConvSubst}[2]{#1\HyphConvFirst}

 \newcommand{\var}[1]{{\text{\rmfamily\slshape\mdseries\HyphConv{#1}\/}}}

 \newcommand{\bd}{\mathbf{b}}
 \newcommand{\G}{\mathbf{G}}
 \newcommand{\Q}{Q}
 \newcommand{\R}{R}

 \newcommand{\maj}{\mathop{\operator@font Maj}}

 \newcommand{\spn}{\var{span}}
 \newcommand{\Span}{\var{Span}}
 \newcommand{\dfl}{\var{defl}}

\makeatother

\begin{document}


\setcounter{figure}{0}

 \newcommand{\vertex}{\circle{0.8}}
 \newcommand{\Vertex}{\circle*{0.8}}

 \newcommand{\vxbox}{\begin{picture}(0,0)
  \put(0,0)\vertex\end{picture}}
 \newcommand{\fullVertex}{\circle*{0.4}}

 \newcommand{\horLine}{\begin{picture}(6,0) 
  \put(0.4,0){\line(1,0){5.2}} \end{picture}}

 \newcommand{\downLine}{\begin{picture}(0,0)
  \put(0.2,-0.33){\line(3,-5){2.6}}
  \end{picture}}
 \newcommand{\downSide}{\begin{picture}(3,5)
  \put(0.2,4.67){\line(3,-5){2.6}} \end{picture}}

 \newcommand{\neDownLine}{\begin{picture}(0,0)
  \put(0.25,-0.25){\line(1,-1){5.5}}
  \end{picture}}

 \newcommand{\upLine}{\begin{picture}(3,5)
  \put(0.2,0.33){\line(3,5){2.6}} \end{picture}}

 \newcommand{\vertUpLine}{\begin{picture}(6,6)
  \put(0,0.4){\line(0,1){5.2}} \end{picture}}

 \newsavebox{\equilBox}
 \newcommand{\equil}{\usebox{\equilBox}}

 \newcommand{\setScale}[1]{ \setlength{\unitlength}{#1}
  \savebox{\equilBox}(0,0)[b]{\begin{picture}(6,5)(-3,0)
    \put(0,0){\vertex}\put(-3,5){\vertex}\put(3,5){\vertex}
    \put(-3,5){\horLine}\put(-3,5){\downLine}\put(0,0){\upLine}
   \end{picture}}
  }

  \newcommand{\neSelfTri}{\begin{picture}(6,6)
    \put(0,0){\vertex}\put(0,6){\vertex}\put(6,0){\vertex}
    \put(0,0){\horLine}\put(0,0){\vertUpLine}\put(0,6){\neDownLine}
   \end{picture}}
\tracingstats=2


 \newcommand{\figOneOne}{\begin{picture}(20,20)
   \multiput(0,6)(6,0){3}{\neSelfTri}
   \multiput(0,0)(6,0){3}{\neSelfTri}
   \put(6,6){\Vertex} \put(7,7){$e_1$} 
   \put(12,6){\Vertex} \put(13,7){$e_2$}
   \put(6,12){\Vertex} \put(7,13){$e_3$}
  \end{picture}}

 \newcommand{\figOneTwo}{\begin{picture}(0,0)
   \multiput(-3,5)(3,5){3}{\equil}
   \multiput(0,0)(3,5){3}{\equil}
   \put(0,10){\Vertex}
   \put(0,12.5){\makebox(0,0){$e_1$}}
   \put(3,15){\Vertex}
   \put(3,17.5){\makebox(0,0){$e_2$}}
   \put(-3,15){\Vertex}
   \put(-3,17.5){\makebox(0,0){$e_3$}} 
  \end{picture}}

\newcommand{\figOne}{\setScale{8pt}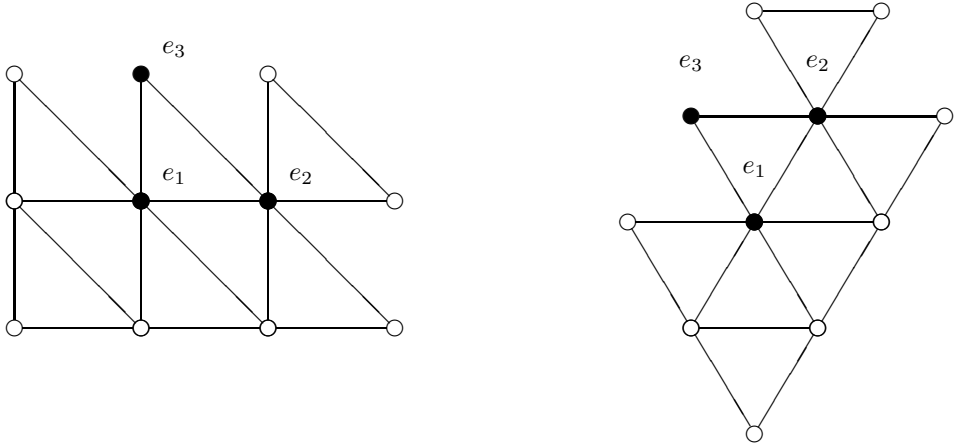
\begin{figure}\centering
   \begin{picture}(50,20)
     \put(0,5){\figOneOne}
     \put(35,0){\figOneTwo}
   \end{picture}
  \caption{ The graph $\G$ and a tile, drawn in the
original and in the symmetrical fashion.}
 \end{figure}}

\newcommand{\figTwo}{\setScale{10pt}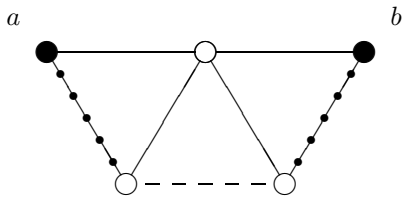
\begin{figure} \centering
 \begin{picture}(12,10)
     \multiput(3,0)(6,0){2}{\equil}
     \multiput(0,5)(12,0){2}{\Vertex}
     \put(-1,6){\makebox(0,0)[rb]{$a$}}
     \put(13,6){\makebox(0,0)[lb]{$b$}}
     \multiput(0.5,4.17)(0.5,-0.83){5}{\circle*{0.3}}
     \multiput(9.5,0.83)(0.5, 0.83){5}{\circle*{0.3}}
     \multiput(3.75,0)(1,0){5}{\line(1,0){0.5}} 
    \end{picture} 
 \label{f:ToomCompon}
  \caption{ To the proof of Lemma~\protect{\ref{l:ToomCompon}} (b).}
 \end{figure}}


 \newcommand{\figThreeOne}{\begin{picture}(0,0)
   \multiput(-6,10)(6,0){3}{\equil} \multiput(18,10)(6,0){3}{\equil} 
   \multiput(-3,5)(6,0){2}{\equil}  \multiput(21,5)(6,0){2}{\equil}
   \multiput(0,0)(24,0){2}{\equil}
   \put(9,15){\horLine}
  \end{picture}}

 \newcommand{\figThreeTwo}{\begin{picture}(0,0)
   \multiput(-3,5)(6,0){2}{\equil}  \multiput(21,5)(6,0){2}{\equil}
   \multiput(0,0)(24,0){2}{\equil}
   \put(12,10){\vertex}
   \multiput(6,10)(6,0){2}{\horLine}
  \end{picture}}

 \newcommand{\figThreeThree}{\begin{picture}(0,0)
   \multiput(0,0)(24,0){2}{\equil}
   \multiput(9,5)(6,0){2}{\vertex}
   \multiput(3,5)(6,0){3}{\horLine}
  \end{picture}}

 \newcommand{\figThreeFour}{\begin{picture}(0,0)
   \multiput(-3,5)(6,0){6}{\equil}  
   \multiput(0,0)(24,0){2}{\equil}
   \multiput(3,5)(6,0){3}{\horLine}
  \end{picture}}

\newcommand{\figThree}{\setScale{4pt}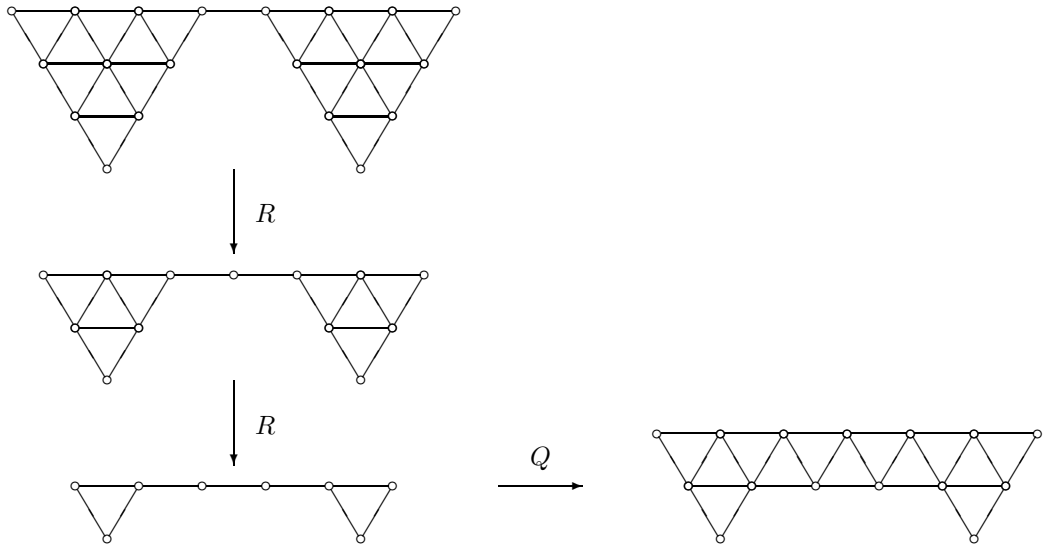
\begin{figure}\centering
   \begin{picture}(100,50)
    \put(9,35){\figThreeOne}
    \put(21,35){\vector(0,-1){8}}
    \put(23,30){$R$}
    \put(9,15){\figThreeTwo}
    \put(21,15){\vector(0,-1){8}}
    \put(23,10){$R$}
    \put(9,0){\figThreeThree}
    \put(46,5){\vector(1,0){8}}
    \put(49,7){$Q$}
    \put(67,0){\figThreeFour}
   \end{picture}
 \label{f:mainTh}
\\ \caption{ The rule $\R^+$ increases thickness.}
\end{figure}}


 \newcommand{\figFourOne}{\begin{picture}(60,50)
   \multiput(15,45)(6,0){5}{\equil}    
   \multiput(15,45)(18,0){2}{\horLine}
   \multiput(24,40)(6,0){2}{\equil}
   \put(27,35){\equil}    
   \multiput(12,40)(-3,-5){4}{\equil}  
   \multiput(0,30)(3,5){4}{\upLine}
   \multiput(42,40)(3,-5){4}{\equil}   
   \multiput(42,50)(3,-5){4}{\downLine}   
   \multiput(6,20)(3,-5){4}{\equil}    
   \multiput(6,30)(9,-15){2}{\downLine}
   \multiput(12,20)(3,-5){2}{\equil}
   \put(18,20){\equil}
   \multiput(48,20)(-3,-5){4}{\equil}  
   \multiput(36,10)(9,15){2}{\upLine}
   \multiput(39,15)(3,5){2}{\equil}
   \put(36,20){\equil}
   \multiput(21,5)(6,0){3}{\equil}     
   \multiput(15,5)(6,0){4}{\horLine}
   \put(27,25){\equil}				       
   \multiput(21,25)(3,5){2}{\upLine}
   \multiput(21,25)(6,0){2}{\horLine}
   \multiput(27,35)(3,-5){2}{\downLine}
  \end{picture}}

\newtoks\tempRow

 \makeatletter 
\def\my@noop#1#2{}
\def\my@until#1#2#3{\if#2#3\let\@nextuntil=\my@noop\else
 #1#3\let\@nextuntil=\my@until\fi\@nextuntil#1#2}

 \chardef\slope=`\\
  \newcommand{\translTok}[1]{\if#1 \relax\else
       \if#1o\vxbox\else
        \if#1-\horLine\else
         \if#1/\upLine\else
          \if#1\slope\downSide\else
           \if#1.\makebox(3,0){}\fi
          \fi
         \fi
        \fi  
       \fi
      \fi}
 \newcommand{\putrow}[2]{\put(#1,#2){\makebox(60,5)[lb]{\expandafter
 \my@until\expandafter\translTok\expandafter>\the\tempRow}}}
  \makeatother

 \def\mv{\catcode`\\=12\global\tempRow=} 

 \newsavebox{\figFourTwo}
 \setScale{3.5pt}
\savebox{\figFourTwo}(60,50){\begin{picture}(60,50) 
   {\mv{  . . .o - o - o - o - o - o - o       > }} \putrow{0}{45}
   {\mv{  . . / \ / \ / \ / \ / \ / \ / \      > }} \putrow{0}{40}
   {\mv{  . .o - o - o - o - o - o - o - o     > }} \putrow{0}{40}
   {\mv{  . / \ / . . . . \ / . . . . \ / \    > }} \putrow{0}{35}
   {\mv{  .o - o. . . . . .o. . . . . .o - o   > }} \putrow{0}{35}
   {\mv{  / \ / . . . . . / \ . . . . . \ / \  > }} \putrow{0}{30}
   {\mv{ o - o. . . . . .o - o. . . . . .o - o > }} \putrow{0}{30}
   {\mv{  \ / \ . . . . / \ / \ . . . . / \ /  > }} \putrow{0}{25}
   {\mv{  .o - o. . . .o - o - o. . . .o - o   > }} \putrow{0}{25}
   {\mv{  . \ / \ . . / \ / \ / \ . . / \ /    > }} \putrow{0}{20}
   {\mv{  . .o - o - o - o - o - o - o - o     > }} \putrow{0}{20}
   {\mv{  . . \ / \ / . . . . . . \ / \ /      > }} \putrow{0}{15}
   {\mv{  . . .o - o. . . . . . . .o - o       > }} \putrow{0}{15}
   {\mv{  . . . \ / \ . . . . . . / \ /        > }} \putrow{0}{10}
   {\mv{  . . . .o - o. . . . . .o - o         > }} \putrow{0}{10}
   {\mv{  . . . . \ / \ . . . . / \ /          > }} \putrow{0}{5}
   {\mv{  . . . . .o - o - o - o - o           > }} \putrow{0}{5}
   {\mv{  . . . . . \ / \ / \ / \ /            > }} \putrow{0}{0}
   {\mv{  . . . . . .o - o - o - o             > }} \putrow{0}{0}
 \end{picture}}

\newcommand{\figFour}{\setScale{3.5pt}\begin{figure}\centering
  \begin{picture}(138,55)
   \put(78,5){\box\figFourTwo}
   \put(0,5){\figFourOne}
   \put(60,25){\vector(1,0){12}}
   \put(60,25){\makebox(12,4)[t]{$R$}}
  \end{picture}
  \label{f:ToomCounter}
 \caption{The rule $\R$ may have adverse effect
on thickness.}
 \end{figure}
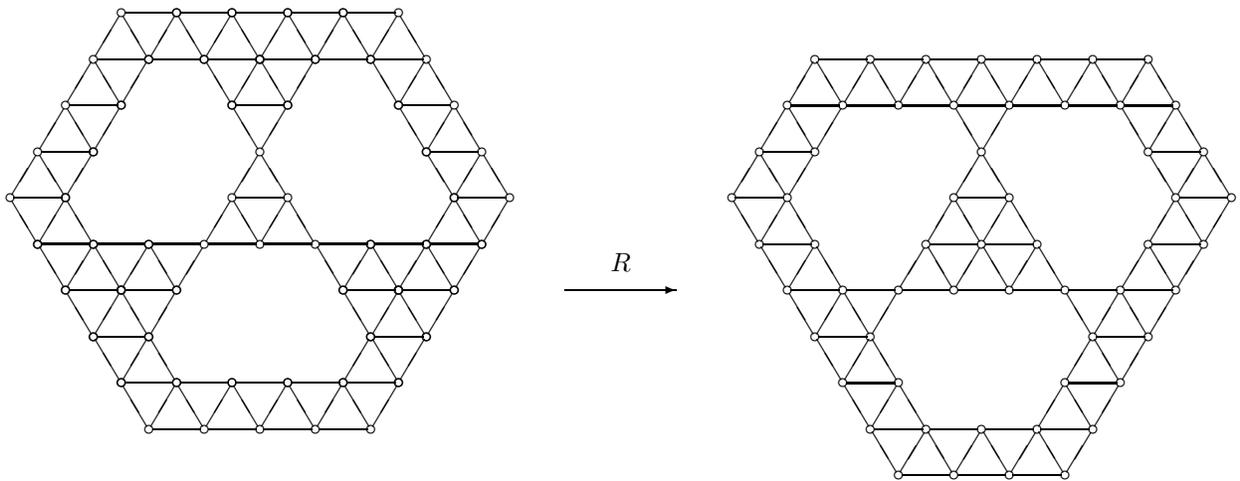}


 \newcommand{\figFive}{\setScale{10pt}\begin{figure}\centering
   \begin{picture}(18,15)
    \multiput(6,10)(6,0){2}{\equil}
    \multiput(3,5)(6,0){3}{\equil}
    \multiput(6,0)(6,0){2}{\equil}
    \put(9,5){\Vertex}
    \put(9,6.5){\makebox(0,0)[b]{$a$}}
    \put(16,13.3){\vector(-1,0){4}}
    \put(17,12.2){\vector(-1,-2){2}}
    \put(18,13.3){\makebox(0,0)[lb]{$P_2(a)$}}
   \end{picture}
\\  \caption{The pairs of tiles with centers in $P_i(a)$.}
 \end{figure}
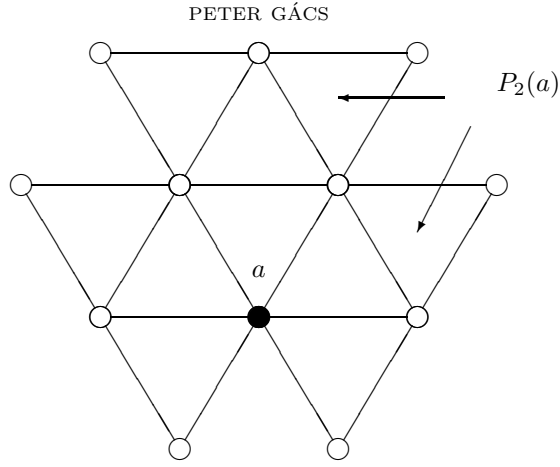}


 \newcommand{\figSix}{\setScale{10pt}\begin{figure}\centering
  \begin{picture}(18,15)
    \put(6,5){\equil}
    \multiput(3,0)(6,0){2}{\equil}
    \put(6,10.6){\makebox(0,0)[b]{$Q(b_1)$}}
    \put(1,2){\makebox(0,0)[rt]{$Q(b_2)$}}
    \put(11,2){\makebox(0,0)[lt]{$Q(b_3)$}}
  \end{picture}
\\  \caption{The tiles $\Q(b_i(t))$.}
 \end{figure}
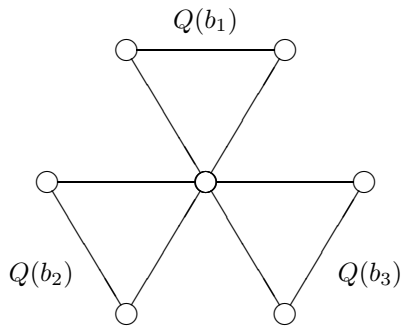}

 \newcommand{\figSeven}{\setScale{10pt}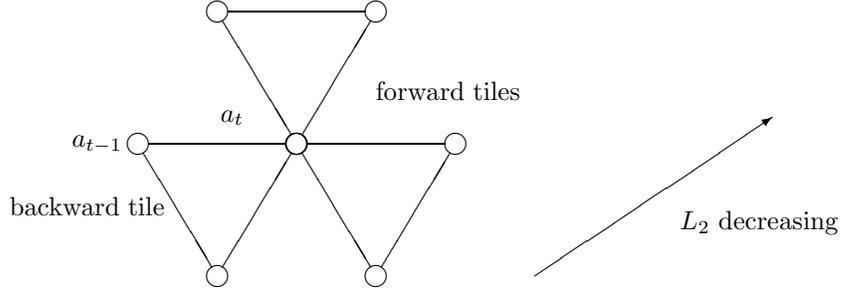
\begin{figure}\centering
  \begin{picture}(18,15)
    \put(6,5){\equil}
    \multiput(3,0)(6,0){2}{\equil}
    \put(4,5.7){\makebox(0,0)[rb]{$a_t$}}
    \put(-0.6,5){\makebox(0,0)[r]{$a_{t-1}$}}
    \put(1,3){\makebox(0,0)[rt]{backward tile}}
    \put(9,7){\makebox(0,0)[l]{forward tiles}}
    \put(15,0){\vector(3,2){9}}
    \put(20.5,2){\makebox(0,0)[l]{$L_2$ decreasing}}
  \end{picture}
\\ \caption{Backward and forward tiles, with $r=2$.}
 \end{figure}}


 \newcommand{\figEight}{\setScale{10pt}\begin{figure}\centering
  \begin{picture}(18,15)
    \multiput(6,5)(6,0){2}{\equil}
    \multiput(3,0)(6,0){3}{\equil}
    \put(6,5){\Vertex}
    \put(4,5.7){\makebox(0,0)[rb]{$a_t$}}
    \put(12,5){\Vertex}
    \put(14,5.7){\makebox(0,0)[lb]{$a_{t+1}$}}
    \put(6,11.6){\vector(0,-1){3.3}}
    \put(6,12.2){\makebox(0,0)[rb]{$Q(b_1(t))$}}
    \put(12,11.6){\vector(0,-1){3.3}}
    \put(12,12.2){\makebox(0,0)[lb]{$Q(b_1(t+1))$}}
   \end{picture}
\\ \caption{To the proof of Lemma \protect\ref{l:last}. The tiles 
$\Q(b_1(t))$ and $\Q(b_1(t+1))$ belong to $S_1$.}
 \end{figure}
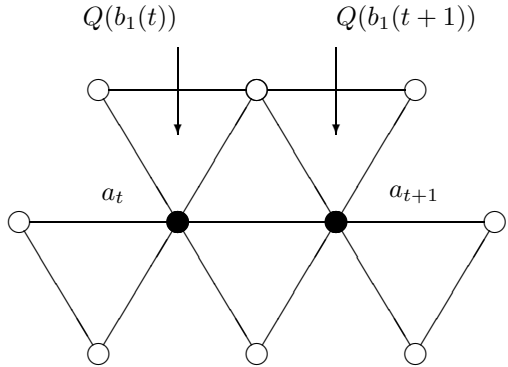}



\date{}
\title{A Toom rule that increases the thickness of sets}

\author{Peter G\'acs}\thanks{Supported in part by NSF grant DCR
8603727.}
\address{Boston University and IBM Almaden Research Center}

\begin{abstract}
 Toom's north-east-self voting cellular automaton rule $\R$ is known
to suppress small minorities.  A variant which we call $\R^+$ is also
known to turn an arbitrary initial configuration into a homogenous
one (without changing the ones that were homogenous to start with). 
Here we show that $\R^+$ always increases a certain property of sets
called thickness.  This result is intended as a step towards a proof
of the fast convergence towards consensus under $\R^+$.  The latter
is observable experimentally, even in the presence of some noise.
 \end{abstract}

 \maketitle

\section {Introduction}

\subsection {Cellular automata}

Cellular automata are useful as models of some physical and
biological phenomena and of computing devices.  To define a cellular
automaton, first a set $\bS$ of possible {\em local states} is
given.  In the present paper, this is the two-element set $\{0,1\}$.
Then, a set $\bW$ of {\em sites} is given.  In the present paper,
this is the two-dimensional integer lattice $\bZ^2$.  A {\em
configuration}, or {\em global state}, $x$ over a subset $B$ of $\bW$
is a function that assigns a state $x[p]\in \bS$ to each element $p$
of $B$.  An {\em evolution} $x[t,p]$ over a time interval
$t_1,\ldots,t_2$ and a set $B$ of sites is a function that assigns a
global state $x[t,\cdot]$ over $B$ to all $t=t_1,\ldots,t_2$.  A {\em
neighborhood} is a finite set $G=\{g_1,\ldots,g_k\}$ of elements of
$\bZ^2$.  A {\em transition rule} is a function $M\such \bS^k\to
\bS$.  An evolution $x[t,p]$ is called a {\em trajectory} of the
transition rule $M$ if the relation
 \begin{equation}
  x[t+1,p]=M(x[t,p+g_1],\ldots,x[t,p+g_k])
 \label{e:traj}\end{equation}
  holds for all $t,p$.  To obtain a trajectory over the whole space
$\bW$, we can start from an arbitrary initial configuration
$x[0,\cdot]$ and apply the local transformation~\eqref{e:traj} to get
the configurations $x[1,\cdot]$, $x[2,\cdot],\ldots$.  The 
rule~\eqref{e:traj} is analogous to a partial differential equation.

Most work done with cellular automata is experimental.  It seems to
follow from the nature of the broader subject (``chaos'') involving
the iteration of transformations that exact results are difficult to
obtain.  The reason seems to be that a trajectory of an arbitrary
transition rule is like an arbitrary computation; and most nontrivial
problems concerning arbitrary computations are undecidable.

Most of the exact work concerns {\em probabilistic} cellular
automata, i.e.\ ones in which the value of the transition rule $M$ is
a probability distribution over $\bS$.  As a simple example, let us
consider a deterministic rule $M$ and an initial configuration
$x[0,\cdot]$.  We begin to apply the relation~\eqref{e:traj} to
compute $x[t+1,p]$ but occasionally, (these occasions occur, say,
independently with a low probability $\rg$), we will violate the rule
and take a different value for $x[t+1,p]$.  The random process
obtained in this way can be called, informally, a {\em
$\rg$-perturbation} of the trajectory obtained from $x[0,\cdot]$.

The most thoroughly
investigated problem concerning probabilistic cellular
automata is a problem analogous to the {\em phase transition} problem
of equilibrium systems (like the Ising model of ferromagnetism).  Given
a probabilistic transition rule, the problem corresponding to the
phase-transition problem of equilibrium systems is whether the
evolution erases all information concerning the initial
configuration.  In that case, it is said that the system does not
have a phase transition.  

The known equilibrium models that exhibit phase transition are not 
known to be stable:  if the parameters are slightly perturbed (e.g.\
an outside magnetic field turned on) the phase transition might
disappear.  In contrast, there are cellular automata exhibiting a
stable phase transition.  It was not a trivial problem to find such
cellular automata.  Indeed, let us look at probabilistic rules
obtained by the perturbation of a deterministic one.  If the rule is
the identity, i.e.\ $x[t+1,p]=x[t,p]$, then this rule remembers the
initial configuration, as long as it is not perturbed.  If it is
perturbed appropriately then the information in the configuration
$x[t,\cdot]$ about $x[0,\cdot]$ converges fast to 0.  Also, most
local majority voting rules seem to lose all information fast when
perturbed appropriately.

\subsection{Toom's rule}

The first rules exhibiting stable phase transition were found by
Andrei Toom.  A general theory of them is given in \cite{Toom80}.

One of Toom's rules is defined with the neighborhood 
 \[
    G=\{(0,0),\ (0,1),\ (1,0)\}
 \]
 and the transition function $M$ which is the majority function
$\maj(x,y,z)$.  In other words, an evolution $x[t,p]$ is a trajectory
of Toom's rule $\R$ if for all $t,p$ where it is applicable, the
following relation holds:
 \[
 x[t+1,p]=\maj(x[t,p], x[t, p+(0,1)], x[t, p+(1,0)]).
 \]
  We will also write 
 \[
   x[t+1,\cdot]=\R(x[t,\cdot]).
 \]
 The rule $\R$ says that to compute the next value in time of
trajectory $x$ at some site we have to compute the majority of the
current values at the site and its northern and eastern neighbors.

For $s=0,1$, let $h_s$ be the homogenous configuration for which
$h_s[p]=s$ for all sites $p$.  The north-east-self voting rule $\R$
is known to suppress small minorities, even in the presence of noise.
If started from a homogenous configuration then the one
bit information saying whether this configuration was $h_0$ or $h_1$,
is preserved.  

There are many variants of the rule $\R$, all of which have the
noise-suppressing property.  One of these was used in
\cite{GacsReif3dim88} to define a simple three-dimensional rule that
can not only store an infinite amount of information about the initial
state but can also simulate the trajectory of an arbitrary
one-dimensional deterministic rule, despite perturbation.

Given the simplicity of the rule $\R$ and its two stable
configurations, it is natural to investigate the effect of repeated
applications of $\R$ to an arbitrary configuration that is close to
neither $h_0$ to $h_1$.  We will identify a configuration $x$ with
the set of sites $a$ where $x[a]=1$.  Therefore we can talk about the
application of $\R$ to a set.  

Let $\G$ be the graph over $\bW$ in which each point is connected to
north, south, east, west, north-west, south-east.  (The graph is
undirected in the sense that with each directed edge, it also
contains the reverse edge.) A subset of $\bW$ is called {\em
connected} if it is connected in $\G$.  Let $S=\bigcup_i S_i$ be a
set with connected components $S_i$. The simple Lemma
\ref{l:ToomCompon} proved later in this paper says that the rule $\R$
does not break up and does not connect the components $S_i$.  For the
plane $\bW=\bZ^2$, the simple Lemma \ref{l:SpanDecr} stated later
says that Toom's Rule ``shrinks'' each of the components, in terms of
the size measure called span.

If the space $\bW$ is the torus $\bZ_n^2$ then the rule $\R$ still
shrinks those connected sets that are isomorphic to subsets of
$\bZ^2$.  These components will be called {\em simple}.  Let us
characterize them.  The {\em increment} of each directed edge
$((a_1,b_1), (a_2,b_2))$ of $\G$ is the vector $(a_2-a_1, b_2-b_1)$. 
The absolute value of both coordinates of this vector is $\le 1$. The
{\em total increment} along a path is the sum of the increments, {\em
without reduction} $\bmod n$.  A closed path (cycle) is {\em simple}
if its total increment is 0.  It is easy to see that a connected
subset of $\bW$ is simple if and only if it does not contain a
non-simple cycle.  Now it is easy to verify the following theorem,
proved in \cite{Gacs2dim89}.  

\begin{theorem} \label{t:ToomLimit}
 Let $S$ be a subset of $\bW$.  The set $\R^i(S)$ becomes eventually
empty as $i\to \infty$ if and only if all components of $S$ are simple.
 \end{theorem}

Thus the minimal sets not erased by the iteration of $R$ are cycles
that wind at least once around the torus.  Toom's rule will not break
up such cycles.  It actually leaves many of them invariant, possibly
shifting them.

\subsection {Global simplification}

There is some interest in trying to find a variant of Toom's rule
that still preserves the stability of the homogenous states $h_s$ but
whose iterations force every configuration $x$ eventually into some
$\bH(x)=h_0$ or $h_1$.  Since there are only two homogenous
configurations, there will be configurations $x,x'$ differing only in
one site, where $\bH(x)=h_0$ and $\bH(x')=h_1$. 

The main interest of such rules comes from the insight they give into
the mechanism of {\em global simplification} of an arbitrary
configuration necessary for such a property.  Of interest is also the
opportunity to investigate the noise-sensitivity of the
simplification, i.e., the size of the attraction domains.

A possible application of such a rule is in situations where a
consensus must be forced from an arbibrary configuration.  The paper
\cite{Gacs2dim89} shows such a situation.  Consensus problems, or, in a
more extravagant terminology, Byzantine Generals Problems, are
central in the area of Computer Science called Distributed
Computing.

\paragraph{Consensus in the absence of failures.}
 Theorem \ref{t:ToomLimit} above suggests a modification of the rule
$\R$ with the desired property. Since the only configurations not
erased by $\R$ are those containing non-simple cycles, we should try
to force all those cycles to $h_1$. This is achieved by biasing the
rule $\R$ slightly in the direction of 1's, while still preserving
the shrinking property given in Lemma \ref{l:SpanDecr}.  We obtain
such a rule $\R^+$ as follows.  To compute the state $\R^+(x)[p]$, of
cell $p$ after applying $\R^+$ to the configuration $x$, apply the
rule $\R$ twice to $x$, then take the maximum of the states of the
neighbors $p$, $p+(0,-1)$, $p+(-1,0)$.  The theorem below shows that
$\R^+$ indeed has the desired limiting consensus property.  Of
course, such a property is interesting only in connection with the
presence of at least two stable configurations.

\begin{theorem} \label{t:Cnsn}
 There is a constant $c$ such that the following holds.
Let $S$ be an arbitrary subset of $\bW=\bZ_n^2$.
Then $(\R^+)^{cn}(S)=h_0$ or $h_1$.
 \end{theorem}

A proof was given in \cite{Gacs2dim89}. Let us sketch here
a more direct proof.  It uses the following lemma from \cite{Gacs2dim89}
saying that the rule $\R^+$ first makes a set fat before erasing it.
The proof is given, for the sake of completeness, in subsection
\ref{ss:Triangles}.

 \begin{lemma} \label{l:Fat}
 Let $S$ be a connected subset of $\bZ^2$ with the property that
$(\R^+)^{2i}(S)\not=\emptyset$.  Then $(\R^+)^i(S)$ has at least
$i^2/2$ elements.
 \end{lemma}

The rule $\R^+$ still has the property of rule $\R$ that it does not
break up connected components.  But, contrary to the rule
$\R$, it can join several components.  The following lemma shows how
the number of components gets smaller, provided no non-simple
component occurs.  (If a non-simple component occurs then the rule
$\R^+$ blows it up anyway, in $\le n$ steps, to occupy the whole
space.)

 \begin{lemma} \label{l:FewerComp}
 Let $C\subset \bW=\bZ_n^2$ have $p$ components, and
$D=(\R^+)^{2i}(C)$ have $q$ components, all of them simple.
If $i\ge n\sqrt{8/p}$ then $q\le 0.75p$.
 \end{lemma}

\begin{proof}
 Let $C_1,\ldots,C_p$ be the components of $C$ and $D_1,\ldots,D_q$
the components of $D$.  Then there is a disjoint union
$\{1,\ldots,p\} = I_1, \cup\cdots\cup I_q$ such that
 \[
   D_j=(\R^+)^{2i}(\bigcup_{k\in I_j} C_k).
 \]
 Let $K$ be the set of those $j$ for which $I_j$ consists of a single
element $i_j$.  These $j$ belong to components $C_{i_j}$ that are
large enough and survive the $2i$ applications of $\R^+$ without
having to merge with other components.  It follows from Lemma
\ref{l:Fat} that $|K|(i^2/2) \le n^2$, i.e., $|K| \le 2(n/i)^2$,
since otherwise, the number of elements of the set
 \[ 
   (\R^+)^i(\bigcup_{j\in K}C_{i_j})
 \]
  would be greater than the number $n^2$ of elements of $\bW$.  Of
course, we have $q-|K|\le p/2$.  Combining these, we have 
 \[
   q \le p/2 + 2(n/i)^2.
 \]
  With $i\ge n\sqrt{8/p}$, we have $q\le 0.75p$.
\end{proof}

\begin {proof}[Proof of Theorem~\protect\ref{t:Cnsn}]
 Let us apply the last lemma repeatedly with
 \[
   C^{k+1} = (\R^+)^{2i_k}(C^k),
 \]
  where $p_k$ is the number of components of $C_k$, and
$i_k=\cei{ n\sqrt{8/p_k}}$.  We get $p_{k+1}\le 0.75 p_k$, 
hence the number of components decreases to 1 fast.  The times
$2i_k$ form, at the same time, approximately a geometric series
in which even the largest term, obtained with $p_k=2$, is at most
$4n$.  Therefore the sum of this series is still $\le c n$ for an
appropriate constant $c$.
 \end{proof}

\paragraph{Consensus in the presence of failures.}
 The sensitivity of the simplification property indicates
difficulties if some violations of the rule are permitted, especially
if these violations are not probabilistic but can be malicious.  It
still follows easily from Theorem \ref{t:Cnsn} that $\R^+$ achieves
near-consensus in $O(n^2)$ steps, even if $o(n)$ of the local
transitions during this procedure were malicious failures. Indeed, in
$\sim n^2$ steps, there is a time interval of size $c n$ with the
constant $c$ of Theorem \ref{t:Cnsn} without failures.  During this
interval, homogeneity is achieved, and given the
stability of the rule $\R^+$, the $o(n)$ failures cannot
overturn it.

Eventually, we would like to show that near-consensus is achieved 
under the same conditions, already in $O(n)$ steps.  This seems true
but difficult to prove.  If failures are permitted the monotonicity
disappears.  Components can not only be joined but also split.  The
argument of Lemma \ref{l:FewerComp} can be summarized thus.
 \begin{quote}
  Small components either disappear or join to survive, therefore
their number decreases fast.  Large components become temporarily fat
therefore their number becomes small.  
 \end{quote}
 If components can also be split then it is possible that small
components join temporarily to survive, then failures split them
again, and thus their number does not decrease.

Hope is given by an observation indicating a property that
is a strengthening of Lemma \ref{l:Fat}.  This lemma says that 
$\R^+$ makes sets fatter before erasing them.  The strengthening
would say that the sets are made not only fat in the sense of
containing many points, but also ``thick'', in the sense of becoming
harder to split.

Informally, a set can be called $k$-thick if for all $i<k$, cutting
off a piece of size $6i$ from it, we need a cutting set of size
approximately $i$. The present paper proves that $\R^+$ indeed has a
thickness-increasing property.  Thus, if $\R^+$ joins two large
components and has $k$ failure-free steps to work on the union then 
the union cannot be split into two large components again by fewer
than $k$ failures.  This application is the informal justification of
the notion of thickness.  

The proof of the thickness-increasing property is a lot of drudgery. 
Its claim to attention rests less on any aesthetic appeal than on
being one of the few examples for the rigorous analysis of an
interesting global behavior of an important cellular automaton.

\section {Some geometrical definitions}

\subsection {Tiles}

Let us call a {\em tile} a triangle $\Q(p)$ consisting of a point $p$
and its northern and eastern neighbors.  Let us call $p$ the {\em
center} of the tile.  We write 
 \begin{align}\label{e:e.i} 
      e_1(p) &= p,\ e_2(p)=p+(1,0),\ e_3(p)=p+(0,1), \nn
    \\ \Q(p) &= \{e_1(p), e_2(p), e_3(p)\}.
 \end{align}
  The ``center'' of the tile is thus really one of the corners.  But it
is better to view the center as identical with the tile itself.
In illustrations, it is better to draw the tiles to be
rotationally symmetric.  The ``center'' of the tile is then the site
at its bottom.  

\figOne

  If the set $S$ intersects a tile in at least two points then we say
that it {\em holds} the tile.  The set $\R(S)$ contains a point $p$
iff $S$ holds the tile with center $p$.  We say that two tiles are
{\em neighbors} if they intersect, or, equivalently, if their centers
are neighbors.  As mentioned above, it is convenient to think of the
graph of tiles instead of the centers themselves, identifying the set
$\R(S)$ with the set of those tiles held by the elements of $S$.  Let 
 \[ 
 \Q(E)=\bigcup_{a\in E} \Q(a).
 \]

\subsection{Triangles} \label{ss:Triangles}

Let us define the linear functions
 \[
 L_1(\ag,\bg)= -\ag,\ L_2(\ag,\bg)=-\bg,\ L_3(\ag,\bg)= \ag+\bg.
 \]
 The triangle $L(a,b,c)$ is defined as follows:
 \[
 L(a,b,c)=\setof{ p\such L_1(p)\le a,\ L_2(p)\le b,\ L_3(p)\le c}.
 \]
 The deflation of the triangle $I=L(a,b,c)$ by the amount $d$ is
defined as follows:
 \[ 
 D(I,d)=L(a-d,b-d,c-d).
 \]
  The {\em span} of the above triangle is the length of its base,
and is given by
 \[
 \spn(I)=a+b+c.
 \]
  For a set $\cI$ of triangles we have
 \begin{align*}
      D(\cI,d) &= \setof{ D(I,d)\such I\in\cI },
 \\ \spn(\cI) &= \sum_{I\in\cI} \spn(I).
 \end{align*}
 For a set $E$ of lattice points, let $\spn(E,d)$ be $\min\spn(\cI)$
where the minimum is taken over sets $\cI$ of triangles covering $E$
with their $d$-deflation, i.e.\ for which $E\sbs\bigcup D(\cI,d)$.  
Here, 
 \[ \bigcup D(\cI,d) = \bigcup_{I\in\cI} D(I,d).
 \]
  Let
 \[ 
 \dfl=2.
 \]
 We write
 \begin{align*}
    \spn(E)   &= \spn(E,0),
  \\  \Span(E) &= \spn(E,\dfl).
 \end{align*}

 The following lemma is easy to verify.

\begin{lemma} \label{l:SpanDecr}
 For a connected set $E$ of lattice points, let $\spn(E,d) > 0$.
 Then 
  \[
   \spn(\R(E),d) = \spn(E,d)-1,\ \spn(\Q(E),d)=\spn(E,d)+1.
  \]
 \end{lemma}

 The number $\spn(E,1/3)$ will be called the {\em discrete span of
$E$}.  The discrete span of a single point is 1.  Two points are
neighbors in $\G$ iff the discrete span of their pair is $\le 2$,
i.e.\ iff the triangles of size 1 around them intersect.  The
following lemma is easy to verify.

 \begin{lemma} \label{l:Addit}
  \begin{itemize}
   \item If two triangles $I_1,I_2$ intersect then
there is a trangle $I$ of size $\spn(I_1)+\spn(I_2)$
containing $I_1\cup I_2$.
   \item If two sets $A_1,A_2$ have neighboring points and
$A_j$ is contained in $D(I_j,1/3)$ for triangles $I_j$ then
there is a trangle $I$ of size $\spn(I_1)+\spn(I_2)$
such that $A_1\cup A_2$ is contained in $D(I,1/3)$.
  \end{itemize}
 \end{lemma}

\begin {proof}[Proof of Lemma \protect\ref{l:Fat}]
 Let $S$ be a connected subset of $\bZ^2$ with the property that
$(\R^+)^{2i}(S)$ is not empty.  We have to give a lower bound on
the set $(\R^+)^i(S)$.

It is easy to verify the following commutation property of the rules
$\R$ and $\Q$:
 \[
   \Q\R(S) \sbs \R\Q(S).
 \]
  It follows that
 \[
  (\R^+)^i(S)\sbs \Q^i\R^{2i}(S).
 \]
  If $(\R^+)^{2i}(S)$ is not empty then $\spn(S,1/3)> 2i$.
It follows from Lemma \ref{l:SpanDecr} that $\R^{2i}(S)$ is not
empty.  The set $\Q^i(\R^{2i}(S))$ then contains a full triangle of
span $i$, which contains $(i+1)(i+2)/2$ elements.
 \end{proof}

The following lemma was used in the definition

\section{The main result}

\subsection {The effect of Toom's rule on components}

Suppose that the set $S$ consists of the connected components
$S_1,\ldots,S_n$.  Connectedness is understood here in the graph
$\G$.  The next statement shows that Toom's rule does not break up or
connect components.  More precisely, it implies that the components
of $\R(S)$ are the nonempty ones among the sets $\R(S_i)$.	 This
statement will not be used directly but is useful for getting some
feeling for the way Toom's rule acts.

\begin{fact} \label{l:ToomCompon} Let $S$ be a subset of $\bW$.
 \begin{description}
 \item[(a)] If $S$ is connected then $\R(S)$ is connected or empty.
 \item[(b)] If $E$ is a connected subset of $\R(S)$ then 
$S\cap\Q(E)$ is connected.
 \end{description}
 \end{fact}

\begin{proof}  
 Proof of (a).  Let $a$ and $b$ be two points in $\R(S)$. Let $a_1$
be a point of $S$ in $\Q(a)$, and $b_1$ a point of $S$ in $\Q(b)$. 
These points are connected in $S$ by a path.  Each edge of the path
is contained in exactly one tile held by $S$.  We have obtained a
path of tiles connecting the tile with center $a$ to the tile with
center $b$.  The centers of these tiles form a path connecting $a$
and $b$ in $\R(S)$.

Proof of (b).  Let $a,b$ be two points in $S_0=S\cap\Q(E)$.  We have
to find a path in $S_0$ connecting them.  Since the set $E$ is
connected it is enough to find such a path when $a,b$ are in two
neighboring tiles, and then work step-by-step.  If the intersection
point of the two neighbor tiles is in $S_0$ then $a,b$ are clearly
connected through it.  Otherwise, $S_0$ contains the edge in both
tiles opposite the intersection.  It is easy to see from 
Figure~\ref{f:ToomCompon} that these two edges have an edge of $\G$
connecting them.
\end{proof} 

\figTwo

\subsection {Cuts and thickness}

For a subset $A$ of $S$, let $\bd_S(A)$ be the set of all elements of
$S\xcpt A$ that are neighbors of an element of $A$.

The triple $(C,A_1,A_2)$ of disjoint subsets of a connected set $S$
in $\bW$ is called a {\em cut of $S$ with parameters $|C|,m$}
if every path in $S$ from $A_1$ to $A_2$ passes through an element of 
$C$, and
 \[ 
 m = \min_{j=1,2} \Span (A_j\cup C).
 \]
  If there is no path from $A_1$ to $A_2$ then $(\emptyset,A_1,A_2)$ 
is a cut.  The cut is called {\em closed} if
 \[ 
    ( \bd_S(A_1) \cup \bd_S(A_2) ) \sbs C.
 \]
  Generally, our constructions will yield a cut $(C,A_1,A_2)$ that is
not necessarily closed.  It can be made closed by adding to $A_j$ all
elements of $S$ reachable from $A_j$ on paths without passing through
$C$.  This operation does not increase the cutting set but increases
the sets $A_j$.  A cut is {\em connected} if both sets $A_j\cup C$
for $j=1,2$ are connected.

Let $\Thg(S,\ag)$ (the {\em $\ag$-thickness} of $S$) denote the
smallest number $k$ such that $S$ has a (not necessarily connected)
cut with parameters $k,m$ with $m > \ag k$.  If no such $k$ exists
then the $\ag$-thickness is $\infty$.  If the $\ag$-thickness
of a large set $S$ is $k$ then a set of cardinality $<k$ cannot cut
off from $S$ a subset of span $> \ag k$, i.e.\ the set $S$ does not
have large parts connected to the main body only on thin bridges.
The main result is the following theorem, showing that the rule 
 \[ 
    \R^+ = \Q \circ \R^2
 \]
 increases the thickness.

\begin{theorem}[Main theorem]\label{t:biasthick}
 We have
 \[ 
    \Thg(\R^+(S),6) \ge \Thg(S,6) +1.
 \]
 \end{theorem}
 As an example let us look at the set on Figure~\ref{f:mainTh}
before and after the application of the rule $\R^+$.  The narrow
connection between the two parts became wider.

\figThree

\subsection {Auxiliary notions of thickness}

The rule $\R$ itself does not increase the thickness of a set.
It cannot even be said that the thickness is preserved.  Though
connections between large parts of the set do not seem to become
narrower, some of these parts may become larger, as the example in
Figure~\ref{f:ToomCounter} shows.  In this example, the three thin
connections holding the central reversed triangle did not become
thicker, but this reversed triangle became bigger.

\figFour

To take these adverse effects into account we need an auxiliary
notion.  Let $\thg(S,\ag,\bg)$ (the {\em $(\ag,\bg)$-thickness} of
$S$) denote the smallest number $k$ such that $S$ has a connected cut
with parameters $k,m$ with $m > \ag k + \bg$.  
Notice that the difference is not only in the extra argument $\bg$
but in that it deals only with connected cuts.  Its relation to
$\Thg(S,\ag)$ is shown by the following theorem.

 \begin{theorem} \label{t:Thg}
 \[
     \Thg(S,\ag) = \thg(S, \ag, 0).
 \]
 \end{theorem}

Before proving this theorem, we need the following lemma.

\begin{lemma} \label{l:eachComp} 
 Let $(C,A,B)$ be a closed cut of $S$ with $|C|< \thg(S,\ag,\bg)$. 
Let us break $A\cup C$ into components $U_1,U_2,\ldots$,
and $B\cup C$ similarly into components $V_1,V_2,\ldots$.  
Then we have either 
 \[
    \Span(U_i)\le \ag |U_i\cap C| + \bg
 \]
 for all $i$, or
 \[
    \Span(V_j)\le \ag |V_j\cap C| + \bg
 \]
  for all $j$.
 \end{lemma} 

\begin{proof}  
 Suppose that the first relation does not hold.  Without loss of
generality, let us assume that
 \begin{equation}
      \Span(U_1) > \ag |U_1\cap C|+\bg.
 \label{e:U.1} \end{equation}
 Let $j$ be arbitrary.  Let $C'=U_1\cap V_j$. Then $C'\subset C$. Let
$A'=U_1\xcpt C'$, $B'=V_j\xcpt C'$.  

The triple $(C',A',B')$ is a connected closed cut of $S$.  The
connectedness follows immediately from the definition.  To show that
it is a closed cut, we have to show $\bd_S(A')\subset C'$.  The
relation $A'\subset A$ implies $\bd_S(A')\subset A\cup
\bd_S(A)\subset A\cup C$, and hence, since $A'\cup C'$ is a component
of $A\cup C$, we have $\bd_S(A')\sbs C'$.

It follows from the fact that $(C',A',B')$ is a connected cut
and from $\thg(S,\ag,\bg) > |C|$ that
 \[ 
    \min (\Span(U_1), \Span(V_j)) \le \ag |C'| +\bg.
 \]
 This, together with~\eqref{e:U.1}, implies
 \[
    \Span(V_j)\le \ag |V_j\cap C| +\bg.
 \]
\end{proof} 

\begin{proof}[Proof of Theorem \protect\ref{t:Thg}]
 Let $S$ be a set with $\thg(S,\ag,0)>k$.  We will estimate
$\Thg(S,\ag)$.  Let $(C,A,B)$ be a closed cut of $S$ with $|C|=k$.
Let us break $A\cup C$ into components $U_1,U_2,\ldots$, and $B\cup
C$ similarly into components $V_1,V_2,\ldots$.  Then, lemma
\ref{l:eachComp} says that we have either 
 \begin{equation}
      \Span(U_i)\le \ag |U_i\cap C|
 \label{e:U.i0} \end{equation}
 for all $i$, or
 \[
     \Span(V_j)\le \ag |V_j\cap C|
 \]
  for all $j$.  Without loss of generality, assume that~\eqref{e:U.i0}
holds.  Then we have
 \[
     \Span(A\cup C) \le \sum_i \Span(U_i) \le \ag|C|.
 \]
 \end{proof} 

When $\bg>0$ then the relation between our notion of $\tg$ thickness
defined (for technical reasons to become clear later) with connected
cuts and a notion defined with arbitrary cuts is not as simple as
above.  The reason can be seen from the last summation in the above
proof.  If we had $\ag|U_i\cap C|+\bg$ instead of $\ag|U_i\cap C|$
then the summation would bring in $n\bg$ where $n$ is the number of
terms.

\subsection {Outline of the proof of the main theorem}

The following theorem, to be proved later, shows that the original
Toom rule ``almost'' preserves thickness.

\begin{theorem} \label{t:Toomthick} 
 If $\bg\le 3\dfl-3$ then
 \[
    \thg(\R(S), \ag, \bg+2) \ge \thg(S,\ag,\bg).
 \]
 \end{theorem} 

The following theorem, to be proved later, says that the rule $\Q$
increases thickness.

\begin{theorem} \label{t:InflThick}
 Suppose that $\bg\le 3\dfl-2$.  Then 
 \[
     \thg(\Q(S),\ag,\bg+2-\ag) \ge \thg(S,\ag,\bg)+1.
 \]
 \end{theorem}

 \begin{proof}[Proof of Theorem \protect\ref{t:biasthick}]
  We apply the above theorems to $\R$,$\R$ and $\Q$ consecutively,
with $\ag=6$ throughout, but with $\bg=0,2,4$ in the three stages.
 \end{proof}

\section {The effect of Toom's rule on thickness}

\begin{proof}[Proof of Theorem \protect\ref{t:Toomthick}]
  Let $U=\R(S)$.  Let $(C,A_1,A_2)$ be a connected cut of $U$ with
$|C|<k$. Without loss of generality, we can assume that it is a
closed cut. Our goal is to estimate $\min_{j=1,2} \Span(A_j\cup C)$. 
We will find a certain cut $(C', B_1, B_2)$ of $S$.

For each element $a$ of $C$, we define an element $a'$ in $S\cap
\Q(a)$, and set $C' = \setof{ a'\such a\in C}$.  To define $a'$, 
remember the notation $e_i$ from~\eqref{e:e.i}.  Let us
group the neighbors of $a$ in three connected pairs $P_i(a)$
$(i=1,2,3)$ where
 \[
      P_i(a) = \setof{ b \not=a \such  e_i(a)\in \Q(b)}.
 \]
 The pair $P_i(a)$ consists of the centers of those tiles containing
the corner $e_i(a)$.  

\figFive

 For each $i$, the pair $P_i$ may intersect one of the sets $A_j$. It
cannot intersect both since $A_1$ and $A_2$ are separated by $C$.  

\begin{itemize}
 \item Suppose that only one pair, say $P_i$, is intersected by
$A_1$, and $e_i(a)\in S$.  Then let $a'=e_i(a)$.  
 \item Suppose that two pairs are intersected by $A_1$, and the third
one, say $P_i$, is not, and $e_i(a)\in S$.  Then let $a'=e_i(a)$.  
  \item In all other cases, we choose $a'$ arbitrarily from the set
$\Q(a)\cap S$.
 \end{itemize}

Now let
 \[ 
    B_j = (S\cap \Q(A_j))\xcpt C'.
 \]

 \begin{lemma}
 The triple $(C',B_1,B_2)$ is a cut.
 \end{lemma}

\begin{proof} 
 It is enough to prove that if there is a path between some elements
$b_j\in B_j$ for $j=1,2$ then this path passes through an element of
$C'$.  Let $b_1=v_1,v_2,\ldots,v_n=b_2$ be such a path.  For both
$j=1,2$, the element $b_j$ is contained in a tile $\Q(a_j)$ for some
$a_j\in A_j$.  Let $q$ be the last $p$ such that $v_p\in \Q(a)$ for
some $a$ in $A_1$.  Let $\Q(w)$ be the tile containing the pair
$\{v_q,v_{q+1}\}$.  Then $w\in C$, since $(C,A_1,A_2)$ is a closed
cut.  It is easy to see from the definition above that $w'$ is either
$v_q$ or $v_{q+1}$.
\end{proof}

Let us complete the proof of Theorem \ref{t:Toomthick}.  We replace
the cut $(C',B_1,B_2)$ with the closed cut $(C',\ol B_1,\ol B_2)$
where $B_i\sbs \ol B_i$. Let $U_1,\ldots,U_n$ be the components of
$\ol B_1\cup C'$, and $V_1,V_2,\ldots$ the components of $\ol B_2\cup
C'$.  It follows from Lemma \ref{l:eachComp} that either 
 \begin{equation}
    \Span(U_i)\le \ag |U_i\cap C'| + \bg
 \label{e:U.i} \end{equation}
 for all $i$, or
 \[
     \Span(V_j)\le \ag |V_j\cap C'| + \bg
 \]
  for all $j$.  Let us suppose without loss of generality 
that~\eqref{e:U.i} holds.  It follows from the definition of $C'$ and $B_1$
that the tile $\Q(a)$ intersects $B_1\cup C'$ for all $a\in A_1\cup
C$.  Let 
 \begin{align*}
    W_i &= \setof{ a\in A_1\cup C\such \Q(a)\cap U_i\not=0 },
 \\ U'_i &= \Q(W_i).
 \end{align*}
  Then $\bigcup_i W_i=A_1\cup C$, $U_i\sbs U'_i$.  It follows from
the connectedness of $U_i$ that $\spn(U_i,\dfl)=\spn(I_i)$ for a
triangle $I_i$ such that $U_i\sbs D(I_i,\dfl)$.  Then the triangle
$J_i=D(I_i,\dfl-1)$ contains $U'_i$, and the triangle $\R(J_i)$
contains $W_i$.  Let $K_i=D(\R(J_i),-1/3)$, i.e.\ the blowup of
$\R(J_i)$ by $1/3$.

Let us call the sets $W_i$, $W_j$ {\em neighbors} if they either
intersect or have neighboring elements.  It follows from the
connectedness or $\bigcup_iW_i$ that the set $\{ W_1,W_1,\ldots\}$ is
connected under this neighbor relation.  Indeed, we constructed $K_i$
in such a way that $W_i\sbs D(K_i,1/3)$.  Therefore if $W_i$ and
$W_j$ are neighbors then $K_i$ and $K_j$ intersect.  Let us call two
triangles $K_i$, $K_j$ {\em neighbors} if they intersect.  Then from
the fact that the set $\{W_1,W_2,\ldots\}$ is connected under the
neighbor relation, it follows that the set $\{K_1,K_2,\ldots\}$ is
also connected under its neighbor relation.

According to Lemma \ref{l:Addit}, 
if triangles $I,J$ intersect then there is a triangle
containing their union whose span is $\le\spn(I)\cup\spn(J)$.  It
follows that there is a triangle $K$ containing $\bigcup_iK_i$ such
that $\spn(K)\le\sum_i\spn(K_i)$.  As we know,
$\spn(K_i,d)=\spn(J_i)+3d-1$ for any nonnegative $d$.  It follows
from~\eqref{e:U.i} that
 \begin{align*}
    \spn(K_i,1/3) &= \spn(J_i)+1-1 = \spn(I_i)-3(\dfl-1)
    \\             &\le \ag |U_i\cap C'| + \bg -3\dfl+3.
 \end{align*}
 We have therefore
 \begin{align*}
          \spn(K) &\le \ag \sum_i |U_i\cap C'| + n(\bg -3\dfl+3)
 \\                &\le \ag |C'| + n(\bg -3\dfl+3)
 \end{align*}
 Finally,
 \begin{align*}
       \spn(A_1\cup C,\dfl) &\le \spn(K,\dfl-1/3)
 \\                          &\le \ag |C'| + n(\bg -3\dfl+3)+3\dfl-1.
 \\                          &\le \ag|C|+\bg+2,
 \end{align*}
  where we used the assumption $\bg\le 3\dfl-3$ to imply that the
coefficient of $n$ is not positive, therefore we can replace $n$ with
1.
 \end{proof} 

\section {The effect of inflation on thickness}

 \begin {proof}[Proof of Theorem \protect\ref{t:InflThick}]
 For a subset $E$ of $\Q(S)$, let 
 \[
     \Q^{-1}(E,S) = \setof{ a\in S\such \Q(a)\cap E \not= \emptyset }.
 \]
 Suppose that $(C,R_1,R_2)$ is a connected cut of $\Q(S)$ with
$|C|\le \thg(S,\ag,\bg)$.  Without loss of generality, we can assume
that it is a closed cut.  Our goal is to estimate
$\min_{j=1,2}\Span(R_j)$.  From the fact that $R_1,R_2$ are separated
by a cut, it follows that the sets $\Q^{-1}(R_j,S)$ are disjoint.
Let 
  \[
     S_j = \Q^{-1}(R_j,S).
  \]
 \begin{lemma} \label{l:S.j}
  We have 
  \[ 
    R_j \sbs \Q(S_j) \sbs R_j\cup C
  \]
   for $j=1,2$.
 \end{lemma}

 \begin{proof}  The first relation follows immediately from the
definition.  For the second relation, note that 
 \[
  Q(S_j)\sbs R_j \cup \bd_S(R_j)
 \] 
  which is contained in $R_j\cup C$ by the 
closedness of the cut $(C,R_1,R_2)$.
\end{proof}

Now we proceed similarly to the proof of
Theorem \ref{t:Toomthick}.  However, we are trying to make the new
cutting set $C'$ smaller than the old one.

\begin{lemma} \label{l:prevthick}
  Let us use the notation introduced above.  There is an element $x$
of $C$, and a mapping $a\to a'$ defined on $C\xcpt \{x\}$ such that
we have $a\in \Q(a')$, and with
 \[
     C'=\setof{ a'\such a\not=x },\ S'_j=S_j\xcpt C'
 \]
  the triple $(C',S'_1,S'_2)$ is a cut of $S$.
 
 \end{lemma}
 The proof of this lemma is left to the next section.

Now we conclude the proof of Theorem \ref{t:InflThick} analogously to
the end of the proof of Theorem \ref{t:Toomthick}.  Let
$(C',\ol S'_1, \ol S'_2)$ be closed cut such that $S'_j\sbs \ol
S'_j$.  Let $U_1,U_2,\ldots$ be the components of $\ol S'_1\cup C'$, and
$V_1,V_2,\ldots$ the components of $\ol S'_2\cup C'$.  It follows from
Lemma \ref{l:eachComp} that either 
 \begin{equation}
      \Span(U_i)\le \ag |U_i\cap C'| + \bg
 \label{e:U.i2} \end{equation}
 for all $i$, or
 \[
      \Span(V_j)\le \ag |V_j\cap C'| + \bg
 \]
  for all $j$.  Let us suppose without loss of generality 
that~\eqref{e:U.i2} holds.  Let $W_i=\Q(U_i)$.  Let us remember the
superfluous element $x$, and define $W_0=\{x\}$.  It follows from our
construction that
 \[
      R_1\cup C \sbs \bigcup_i W_i.
 \]
 It follows from the connectedness of $U_i$ that $\Span(U_i)=
\spn(U_i,\dfl) = \spn(I_i)$ for a triangle $I_i$ such that $U_i\sbs
J_i=D(I_i,\dfl)$.  Then $W_i\sbs \Q(J_i)$.  Let $K_i=D(\Q(J_i),-1/3)$
for $i>0$, and $D(\{x\},-1/3)$ for $i=0$.  Just as in the proof of
Theorem \ref{t:Toomthick}, we can conclude that there is a triangle
$K$ containing $\bigcup_i K_i$ such that $\spn(K) \le
\sum_i\spn(K_i)$.  It follows from~\eqref{e:U.i2} that, for $i>0$,
 \begin{align*}
     \spn(K_i) &= \spn(J_i)+1+1 = \spn(I_i)-3\dfl+2
 \\                 &\le \ag |U_i\cap C'| + \bg -3\dfl+2.
 \end{align*}
 We have therefore
 \begin{align*}
     \spn(K) &\le \ag \sum_i |U_i\cap C'| + n(\bg -3\dfl+2)
  + \spn(K_0)
  \\          &\le \ag |C'| + n(\bg -3\dfl+2)+1.
 \end{align*}
 Finally,
 \begin{align*}
  \spn(R_1\cup C,\dfl) &\le \spn(K,\dfl-1/3)
 \\                    &\le \ag |C'| + n(\bg -3\dfl+2)+1+3\dfl-1
 \\                   &\le \ag(|C|-1)+\bg+2 = \ag |C| + \bg + 2 -\ag
 \end{align*}
  where we used the assumption $\bg\le 3\dfl-2$ to imply that the
coefficient of $n$ is not positive, therefore we can replace $n$ with
1.
 \end{proof} 

\section {Cutting the pre-image with fewer points}

\subsection {Conditions for a cut in the pre-image}

\begin{proof}[Proof of Lemma \protect\ref{l:prevthick}]
 In later parts of the proof, we will give an algorithm for the
definition of the distinct elements $a_1,a_2,\ldots$, the number
$s$ with $a_s=x$, and the sets
 \[
     C'_t =\setof{ a'_i\such i\le t,\ i\not=s }.
 \]
 Let $S_i^t=S_i\xcpt C'_t$.  Let $C'_0=\emptyset$.  Assume that
$a_1,\ldots,a_t$ and $C'_{t-1}$ have already been defined.  First we
see that, given $a_1,a_2,\ldots,a_t$, what conditions must be
satisfied by $s$ and $a'_t$ to make $(C',S'_1,S'_2)$ a cut of $S$.

The element $a_t$ is contained in three tiles $\Q(b_i(t))$ for
$i=1,2,3$.  They are numbered in such a way that 
 \[
     a_t=e_i(b_i(t)).
 \]
  Let us write $B(t)=\{b_1(t),b_2(t),b_3(t)\}$.

\figSix

 We say that $a_t$ is {\em superfluous} if one of the $S_j^{t-1}$
does not intersect the set $B(t)$.  We will choose $a_1,a_2,\ldots$
later in such a way that there is a $t$ such that $a_t$ is
superfluous.  

 \begin{condition} \label{c:x}
  The point $a_s$ is the first superfluous element of the sequence
$a_1,a_2,\ldots$.
 \end{condition}

If $a_t$ is not superfluous then there is a $b$ and $j$ such that 
 \[
     \{b\}=B(t)\cap S_j^{t-1}.
 \]
  Such a $b$ is called {\em eligible} for $t$.  Let $E(t)$ be the set
of those (one or two) elements of $B(t)$ that are eligible for $t$. 

 \begin{condition} \label{c:elig}
  If $a_t$ is not superfluous then $a'_t\in E(t)$.
 \end{condition}

\begin{lemma} \label{l:elig}
 If conditions \ref{c:x} and \ref{c:elig} are satisfied
then $(C',S'_1,S'_2)$ is a cut of $S$.
 \end{lemma}

\begin{proof} 
 Suppose that there is a path $u_1,\ldots,u_n$ going from $S'_1$ to
$S'_2$ in $S$.  Let $u_p$ be the first element of the path that is
not in $S'_1$.  We will prove that it is in $C'$.  The point $a$ in
the intersection of $\Q(u_{p-1})$ and $\Q(u_p)$ is the neighbor of an
element of $R_1$, since it is in $\Q(u_{p-1})$.  If it is an element
of $R_1$ itself then $u_p\in S_1$.  Since $u_p\not\in S'_1$, it
follows that $u_p\in C'$ and we are done.

Suppose therefore that $a\not\in R_1$.  Then $a\in C$, since
$(C,R_1,R_2)$ is a closed cut.  Let $t$ be such that $a=a_t$.  Then
$u_{p-1}\in S_1^{t-1}$.  If $u_p\not\in C'_{t-1}$ then $u_p\in
S_2^{t-1}$, by the definition of $S_2$.  Then $a_t$ is not
superfluous, and by Condition \ref{c:elig}, $a'_t$ is either
$u_{p-1}$ or $u_p$.
\end{proof}

\subsection {The choice of $a'_t$ and $a_{t+1}$}

After Lemma \ref{l:elig}, what is left from Lemma \ref{l:prevthick}
to prove is that the sequences $a_t,a'_t$ can be chosen satisfying
Condition \ref{c:elig} in such a way that one of the $a_t$ is
superfluous.

The construction will contain an appropriately chosen constant
$r=1,2$ or 3.  If 
  \begin{equation}
   a_{t-1}\in \Q(b_r(t))
  \label{e:rIncr} \end{equation}
 then we say that a {\em forward choice} is made at time $t$.  In
this case, $a_t$ is in corner $r$ of the tile containing both $a_t$
and $a_{t-1}$.  We call this tile the {\em backward} tile.  The
value of the linear function $L_r$ is greater on $a_{t-1}$ than on
$a_t$. Let us call the two other tiles containing $a_t$ the {\em
forward} tiles.  

\figSeven

 The set
 \[
      F(t)= B(t)\cap (S_1^{t-1}\cup S_2^{t-1}) \xcpt \{ b_r \}
 \]
  is the set of the centers of one or two forward tiles for $t$.  In
case of a forward choice, the corner $r$ of one of the forward tiles
is chosen for $a_{t+1}$.  
Suppose that there is a $b$ in $F(t)$ satisfying 
 \begin{equation}
      e_r(b)\in C\xcpt\{a_1,\ldots,a_t\}.
 \label{e:Cavail}\end{equation}
  Then choosing $a_{t+1}$ as such a $b$ would make a strong forward
choice.

If, in addition to~\eqref{e:rIncr}, we also have $a_{t+1}=e_r(a'_t)$
then we say that a {\em strong forward choice} is made.

\begin{condition} \label{c:strongforw}
 Suppose that there is a $b$ in $E(t)\cap F(t)$ 
satisfying~\eqref{e:Cavail}.  Then $a_{t+1}$ is such a $b$, and with $a'_t=b$
a strong forward choice is made.
 \end{condition}

Conditions \ref{c:elig}, \ref{c:strongforw} are the only ones
restricting the choice of $a'_t$ and $a_{t+1}$ for $t>1$.  Otherwise,
the choice is arbitrary.

 \begin{lemma}\label{l:strongforw}
 Suppose that no superfluous $a_i$ was found for
$i=1,\ldots,t$, all earlier choices (if any) were forward, and 
 \begin{equation}
  F(t)\cap S_j^{t-1}\not=\emptyset\ \for j=1,2.
 \label{e:forw} \end{equation}
  Then there is a $b$ in $F(t)$ satisfying~\eqref{e:Cavail} and
therefore a forward choice can be made.
If there is a $b$ in $E(t)\cap F(t)$ satisfying~\eqref{e:Cavail} 
then all choices beginning with $t$ are strongly forward, until a
superfluous node is found.
 \end{lemma}

\begin{proof} 
 By the assumption~\eqref{e:forw}, the elements of $F(t)$ are contained
in two different sets $S_j$.  It follows from Lemma \ref{l:S.j} that
the two forward tiles are contained in different sets $R_j\cup C$.  
There is an edge between the corners $r$ of the two forward tiles. 
Since $C$ separates $R_j$, it must contain one of these points
$e_r(b)$.  Since all our earlier choices were forward, the function
$L_r$ is strictly decreasing on the sequence
$a_1,a_2,\ldots,a_t,e_r(b)$.  Therefore it is not possible that
$e_r(b)$ is equal to one of the earlier elements of the sequence, and
hence~\eqref{e:Cavail} is satisfied.

If a $b$ in $F(t)\cap E(t)$ can be found satisfying~\eqref{e:Cavail}
then according to Condition \ref{c:strongforw}, the strong forward
choice $a'_t=b$, $a_{t+1}=e_r(b)$ is made.  From $a'_t\not\in
S_1^t\cup S_2^t$, it follows that either $a_{t+1}$ is superfluous or
$E(t+1)=F(t+1)=B(t+1)\xcpt \{a'_t\}$.  In the latter case, the
conditions of the present lemma are satisfied for $t+1$, implying
that the next choice is also strong forward, etc.
\end{proof} 

\subsection {The choice of $r,a_1,a'_1$ and $a_2$}

 \begin{condition} \label{c:a.1}
  \begin{enumerate}
   \item If $a_1$ can be chosen superfluous then it is chosen so.
   \item If $a_1$ cannot be chosen superfluous but it can be chosen
to make $|E(1)|>1$ then it is chosen so.  In this case, $r$ is chosen
to make $E(1)=F(1)$.
  \end{enumerate}
 \end{condition}

If the second case of the above condition occurs then all conditions
of Lemma \ref{l:strongforw} are satisfied with $t=1$.

 \begin{condition} \label{c:a.2}
 Suppose that none of the choices of Condition \ref{c:a.1} are
possible, and $r,a_1,a'_1,a_2$ can be chosen to either make $a_2$
superfluous or to satisfy the conditions of Lemma \ref{l:strongforw}
with $t=2$.  Then they are chosen so.
  \end{condition}

\begin{lemma} \label{l:last} 
 The elements $r,a_1,a'_1,a_2$ can always be chosen in such a way that 
either Condition \ref{c:a.1} or Condition \ref{c:a.2} applies.
 \end{lemma}

Before giving the proof of this lemma, let us finish, with its help,
the proof of Lemma \ref{l:prevthick}.  The complete algorithm of
choosing $a_t,a'_t,r$ is as follows.  Choose $a_1$ to satisfy
Condition \ref{c:a.1}.  If the second part applies then choose $r$
accordingly.  If Condition \ref{c:a.2} applies then choose
$r,a'_1,a_2$ to satisfy Conditions \ref{c:elig} and \ref{c:a.2}. From
now on, choose $a'_t,a_{t+1}$ to satisfy Conditions \ref{c:elig} and
\ref{c:strongforw}.

A superfluous $a_t$ will always found.  Indeed, if the first part of
Condition \ref{c:a.1} applies then $a_1$ is superfluous.  If the
second part applies then the conditions of Lemma \ref{l:strongforw}
are satisfied with $t=1$.  If Condition \ref{c:a.2} applies then they
are satisfied with $t=2$.  From this time on, strong forward choices
can be made until a superfluous $a_t$ is found.  This is unavoidable
since $C$ is finite and hence we cannot go on making strong forward
choices forever.
 \end{proof} 

\begin {proof}[Proof of Lemma \protect\ref{l:last}]
 Suppose that the statement of the lemma does not hold.  We will
arrive at a contradiction.  Choose $a_1$ arbitrarily.  We have
$|E(1)|=1$.  We can choose $r$ to get $|F(1)|=2$, $E(1)\sbs
F(1)$.  We will show that we can then make a forward choice (not
strong) for each $t$ and recreate the conditions~\eqref{e:forw}
indefinitely.  This is the desired contradiction since our set is
finite.

Assume that we succeeded until $t$.  By lemma \ref{l:strongforw},
there is a $b$ in $F(t)$ such that~\eqref{e:Cavail} holds.  If $b\in
E(t)$ then with the choice $\ol a_1=a_t$, $\ol a'_1=b$, $\ol
a_2=e_r(b)$ Condition \ref{c:a.2} would apply, and we assumed this is
impossible.  Therefore $b\not\in E(t)$.

Without loss of generality, let us assume 
 \[
      E(t)=\{b_1(t)\}\sbs S_1^{t-1},\ r=2.
 \]

\figEight

  Then $b\not=b_1(t)$.  From $b\in F(t)$, it follows that
$b\not=b_2(t)$, hence $b=b_3(t)$.  Since $a_t$ is not superfluous,
the assumption $E(t)=\{b_1(t)\}$ implies 
 \[
   B(t)\cap S_1^{t-1}=\{b_1(t)\},\ B(t)\cap S_2^{t-1}=\{b_2(t),b_3(t)\}.
 \]
  Let us show $b_1(t+1)\in S_1^{t-1}$.  It is easy to check that the
two tiles $\Q(b_1(t))$ and $\Q(b_1(t+1))$ intersect in $a =
e_2(b_1(t))= e_3(b_1(t+1))$.  If $b_1(t+1)$ belonged to $S_2^{t-1}$
then, by Lemma \ref{l:S.j}, the tile $\Q(b_1(t+1))$ would be
contained in $R_2\cup C$ while for similar reason, the tile
$\Q(b_1(t))$ is contained in $R_1\cup C$.  Then the intersection
point $a$ would have to belong to $C$.  But then we could 
satisfy~\eqref{e:Cavail} with $b=b_1(t)\in E(t)$.  

We have $b_3(t+1)\in S_2^{t-1}$.  Indeed, if it belonged to
$S_1^{t-1}$ then the choice $\ol a_1=a_{t+1}$, $\ol a_2=a_t$ would
again satisfy all conditions of Lemma \ref{l:strongforw} which we
supposed is impossible.  We found that the neighborhood of
$a_{t+1}$ is just a shift of the neighborhood of $a_t$.  This could
continue indefinitely.
 \end{proof} 

\section {Conclusion}

Let us make a remark on the possible extension of the
present work.  The presence of failures seems to necessitate a more
complicated notion of thickness, and it is not clear what
the appropriate generalization of the main theorem should be in that
case.

A variant of the main theorem can probably be proven where the size
of the cutting set is measured in terms of its span instead of number
of elements.  If the proof of that variant is significantly simpler
then it should replace the present theorem.

The stability property of the rules analogous to Toom's rules can
also be proved for continuous-time systems.  In such systems, the
transition rule is not applied simultaneously at all sites, rather
each site applies it at random times.  It seems that the consensus
property of slightly biased Toom rules holds also for this
situation.  Though the methods used in the present paper seem to
depend on synchrony, especially the fact that the inflation
operation is carried out all at once, it is hoped that the concepts
will be useful in extensions to these related problems.

\paragraph{Acknowledgement:}
 The author is thankful to the anonymous referee for the careful
and thorough reading and helpful remarks.

\paragraph{Key words:} Cellular automata, Toom's rule, statisical mechanics.

\end{document}